\documentclass[12pt,a4paper]{article}
\usepackage{graphicx} 
\usepackage{color}
\usepackage{hyperref}
\usepackage{cclicenses}
\usepackage{wrapfig}

\definecolor{green}{rgb}{0.0, 0.4, 0.0}
\newcommand*{\gsd}{\textcolor{green}{\textsf{\textit{Green Software Development}}}}
\newcommand{\concept}[1]{\textsf{\textbf{#1}}}

\title{Estimating the Energy Footprint of Software Systems: a Primer\footnote{This document is distributed with a Creative Commons Attribution-NonCommercial-ShareALike 4.0 license. \byncsa}}
\author{Fernando Castor\\University of Twente}
\date{\today}

\begin{document}  

\maketitle

\section{Introduction}
Software systems do not directly consume energy, since software is just information. These systems are stored on and executed by computing devices, e.g., smartphones, desktop computers, servers, etc., and determine what these devices do and how they do it. The operation of computing devices consumes electrical energy. The bits that computers use to represent information are positive and negative electrical signals. For simplicity, we refer to the energy used by a computing device as a consequence of the instructions of a software system as the energy footprint, or, synonymously, energy consumption or energy usage, of the software system. When talking about the energy footprint of software, we assume that other factors that may impact the energy usage of the underlying computing device can be controlled or reliably accounted for to isolate the impact of the software system.

In \gsd, quantifying the energy footprint of a software system is one of the most basic activities. In the same way that we can verify whether a change to a software system is a performance improvement by measuring the time it takes to perform a certain activity before and after the change, we can measure how much energy is consumed to verify whether a change has made the system more energy efficient. Moreover, if there is a well-established notion or threshold to determine what it means for a system to consume too much energy, e.g., not more than  5\% of the battery of a certain model of mobile device per hour of typical continuous use, measurement can help developers establish whether that limit is respected or not. In this document, we talk about how the energy footprint of a software system can be estimated to support \gsd. Our focus is on general concepts and approaches and not on specific tools, although we do refer to some of them to make explanations more concrete. 

This document is structured as follows. Section~\ref{sec:definitions} introduces the two high-level techniques for performance analysis, with an emphasis on \gsd: measurement and modeling. Section~\ref{sec:ex} then presents a detailed example of how we can conduct an experiment to estimate the energy consumption of a simple benchmark. The section not only explains how the experiment is conducted, but also highlights some considerations in experiment design, execution, and reporting. Section~\ref{sec:approaches} then presents three concrete approaches to estimating software energy consumption: (i) hardware performance counters + software, (ii) specialized measurement hardware, and (iii) analytical models. Section~\ref{sec:tools} makes some observations about readily-available tools. The document wraps up in Section~\ref{sec:further} with suggestions for further study.










\section{Measurement and modeling}\label{sec:definitions}

There are two main techniques to estimating the energy footprint of a software system: measurement and modeling. Different approaches within these techniques have diverse characteristics in terms of \concept{invasiveness}, \concept{accuracy}, and \concept{granularity}, among other attributes~\cite{John:2000:PET}. An ideal approach is non-invasive, i.e., it does not affect what is being evaluated, accurate, i.e., the values it reports perfectly match what is being observed, and supports a wide range of levels of granularity, from whole system all the way to individual lines of code. In addition, it should be easy to use, cheap, and usable either before or after a system is built. In practice, there are no ideal approaches and it is necessary to consider trade-offs.

\subsection{Measurement}\label{sec:measure}

\begin{wrapfigure}{r}{0.45\textwidth}
    \includegraphics[width=1\linewidth]{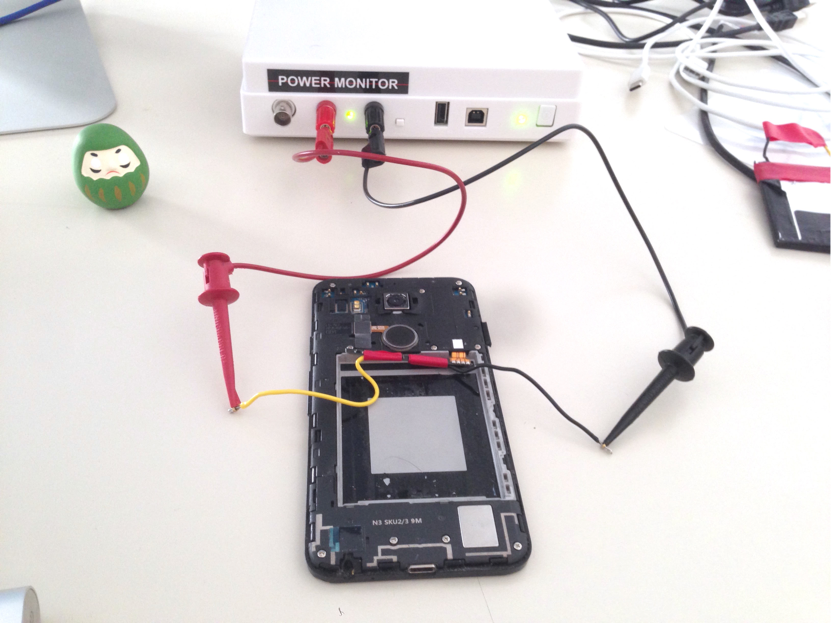}
    \caption{A smartphone connected to a Monsoon power meter. The back case of the phone must be removed to take out the battery. Image by Luis Cruz~\cite{cruz2021tools}.}
    \label{fig:power}
\end{wrapfigure}

As the name implies, measurement entails the use of a \concept{measuring instrument}, e.g., external equipment or features made available by contemporary processors and GPUs, to compute how much energy or related quantity, e.g., power, current, or battery charge, is consumed during the operation of a computing device executing a software system, under a specific workload. Measurement can be performed for the whole device (as shown in Figure~\ref{fig:power}) or for specific components, such as the CPU, GPU, motherboard, etc.  

\noindent
The main positive aspect of measuring energy consumption is that we can directly quantify the attribute of interest (energy). Another positive aspect is that measurement equipment can be completely external to the computing device under measurement. As a consequence, it does not interfere with the the execution of the system and does not create overhead, i.e., it is non-invasive. 
That is the case of power meters. A negative aspect of this approach is that it may be  difficult to connect the results to software elements. This approach usually targets either a whole computing device or specific parts of it, such as the processor. For example, we cannot directly measure how much energy a method, a thread, of a class uses, although it is possible to estimate them~\cite{Couto:2014:DAE,Hao:2013:EMA,Nucci:2017:PSB}. Another disadvantage is that the system or some representative prototype must be available to be measured. This may sound obvious, but in many cases, it is preferable to know the resource requirements of a system before it is (completely) built. Finally, an aspect that can be either positive or negative is that the precision of the measurement is dependent on the sampling rate of the measuring instrument. 
 
Measuring energy footprint can give us readings that are precise enough in practice to make decisions about energy optimization. Nevertheless, even properly functioning measurement equipment may lack sufficient precision to capture the phenomenon of interest appropriately. In any scenario where we convert a continuous-time signal (such as an electrical current) into a discrete-time signal, we must observe the frequency with which values of the continuous-time signal are obtained. This frequency is called the \concept{sampling rate} or \concept{sampling frequency}. It is measured in Hz (or kHz), i.e., number of occurrences (or thousand occurrences) per second. Music presents an illustrative example. To convert live music or music playing from an old vinyl record into digital format, e.g., a CD or an mp3 file, the waves of which the music is comprised must be sampled at regular intervals. For mp3, a typical sampling rate is 48kHz, i.e., 48,000 observations per second. If instead we use a much lower sampling rate such as 4kHz, we may loose relevant information about the music and the resulting digital format song will sound poorly\footnote{This statement does not aim to foment the holy war between analog and digital music formats that music aficionados tend to often get dragged into.}  In a similar vein, when we ``measure'' power, we are actually sampling the power at regular intervals and the resulting energy is just the aggregation of these measurements over time.  The performance counters that keep track of energy usage in Intel processors are updated at a frequency of 1kHz~\cite{Khan:2018:RAE} with information about the energy consumption of the processor and RAM. An industry grade power meter, such as Monsoon\footnote{https://www.msoon.com/high-voltage-power-monitor} has a sampling rate of 5kHz. A high-end digital-to-analog converter\footnote{https://www.ni.com/docs/en-US/bundle/usb-6003-specs/resource/374372a.pdf}, to which we can connect a current clamp to non-intrusively capture current, e.g., from a computer's power source to its motherboard, can have a sampling rate of up to 100 kHz!  

Leveraging measurement equipment with a low sampling rate, such as 1Hz, can have negative consequences. If a program executes too fast, e.g., takes less than 1s to execute, the power meter will not capture its execution. Even if the program runs for a few seconds, the low sampling rate means that only a few observations will be collected. In this scenario, the meter might (i) not capture the entire execution of program, e.g., miss the tail end of the execution, (ii) capture too few observations, e.g., completely ignoring moments where power is high if they happen between subsequent observations. Such low sampling rate equipment may still be useful, though, if execution takes a long time, e.g., one hour. We come back to this topic with an example of how a high sampling rate may also be problematic in Section~\ref{sec:results}.

\subsection{Modeling}

The counterpart to measuring the energy footprint is modeling it. This involves the creation of an analytical or simulation-based, i.e., trace-based, model of the software system or some of its relevant aspects. One advantage of this technique is that it eschews the use of specialized equipment. Furthermore, it can be used in scenarios where the software system cannot be measured, for example, because it has not been built yet or because it is running in an environment over which we have limited or no control, such as the cloud. Yet another point in favor of modeling is that we can obtain estimates for the energy footprint of fine-grained, software-specific elements, such as methods~\cite{Nucci:2017:PSB,LinaresVasquez:2014:MEG}, data structures~\cite{oliveira2021improving,Lima:2019:HEE}, and even lines of code~\cite{Hao:2013:EMA}. These estimates do not even need to directly quantify energy; in some cases, it is sufficient to just enable comparison between alternative solutions~\cite{oliveira2021improving}. In spite of its positive points, the use of models to estimate energy usage for software has a number of downsides. The first and most obvious one is accuracy. Models rely on simplifying assumptions that may reduce their ability to produce accurate estimates. One can leverage real world execution data to create more accurate models. However, in this case the software system or a representative prototype must be available. Furthermore, this often involves some kind of instrumentation, i.e., adding instructions to the system with the sole purpose of producing monitoring data, which is invasive and creates runtime overhead. Furthermore, creating a model is non-trivial. Although some models may be as simple as a linear equation, models for complex systems tend to be expensive to construct and require specialized formalisms, such as petri nets~\cite{Callou:2011:ECE}. 

Models often leverage the results of measurements to make them more realistic. Although it is possible to build a model that is entirely anaytical, e.g., based on low-level specifications of how machine instructions work~\cite{Callou:2011:ECE}, it is common for models to be designed based on real-world measurements or finer-granularity, high-fidelity models, so as to focus on higher-level aspects~\cite{Babakol:2024:TAE,Hao:2013:EMA,oliveira2021improving}. For example, Babakol and Liu~\cite{Babakol:2024:TAE} leverage data available in hardware-specific performance counters (more about this in Section~\ref{sec:counters}) for both CPU and GPU to derive a model that can pinpoint energy consumption hotspots in the layers and abstractions within a deep neural network. 

In some cases, it is possible to use models that estimate performance in terms of execution time as a proxy for energy footprint~\cite{Cruz:2017:PBG,Albers:2012:RIN}. There is convincing evidence that, for applications that are CPU-bound, sequential, and where it is possible to identify specific performance bottlenecks, one metric is an excellent predictor for the other~\cite{Lima:2019:HEE}. This is not true in general, though, for a number of reasons. First, there are aspects of the performance of a system cannot be measured by execution time, e.g., precision of location services, but still consume energy. Second, some approaches to improve performance inherently consume more energy, for example, adding more servers to a data center to reduce request latency~\cite{Dean:2013:TS}. Third, concurrent and parallel software may run faster at the cost of wasteful computations, e.g., due to spinning or transaction retries\cite{Lima:2019:HEE,Gregg:2017:CUW}. Fourth, execution time only partially considers network and disk accesses. A similar case can be made for employing CPU utilization as a proxy for power~\cite{Gregg:2017:CUW,Cockcroft:2006:UVU}. In summary, although performance can be strongly connected to energy usage in software, the two attributes are not synonyms.






\section{Estimating software energy consumption: An illustrative example}\label{sec:ex}

\begin{wrapfigure}{r}{0.50\textwidth}
    \includegraphics[width=1\linewidth]{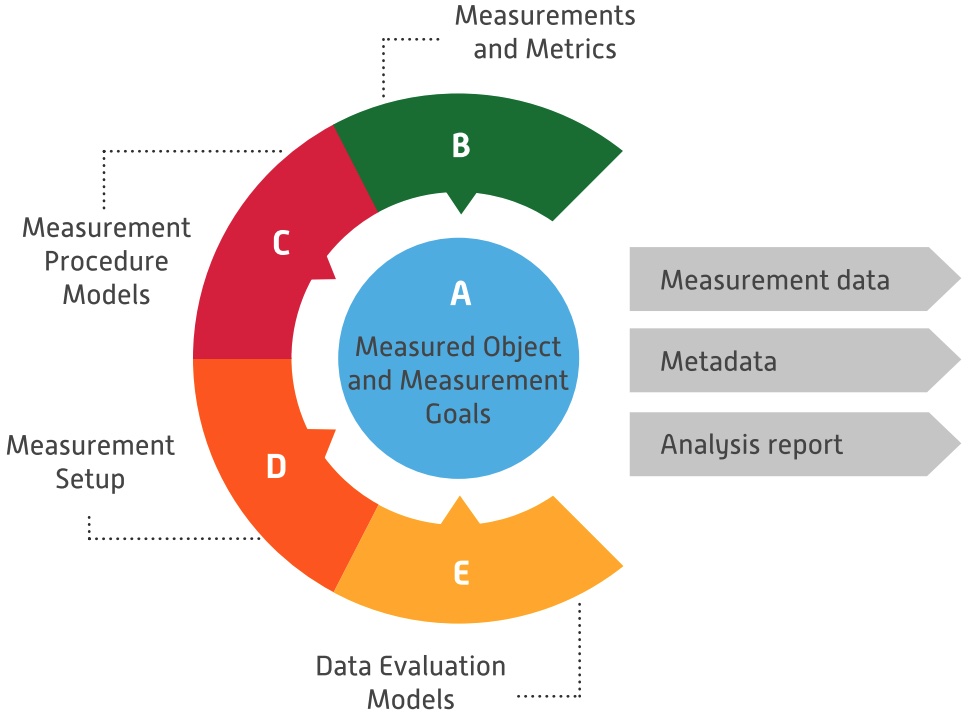}
    \caption{A schematic overview of the measurement process in \gsd. Image by Guldner et al.~\cite{Guldner:2024:DER}}
    \label{fig:schematics}
\end{wrapfigure}

This section provides an example of how to estimate the energy footprint of a small parallel benchmark.  Figure~\ref{fig:schematics} presents a high-level schematic representation of the measurement process in \gsd, taken from the work of Guldner and colleagues~\cite{Guldner:2024:DER}. The benchmark we study in this section is called Fannkuch Redux and it is part of the Computer Languages Benchmark Game\footnote{\url{https://benchmarksgame-team.pages.debian.net/benchmarksgame/}} (CLBG). The latter is a website aimed at comparing the performance of different programming languages. Its benchmarks have been used in a plethora of studies on \gsd~\cite{Lima:2019:HEE,Oliveira:2017:SEC,pereira2021ranking,Pinto:2014:UEB}. 

We attempt to answer a basic research question: \textbf{how much energy does it consume, on average, on the author's personal computer?} The remainder of this section highlights some issues that deserve attention when estimating energy usage and aspects of the environment that should be reported in this kind of study. As a consequence of this approach, the subsections include a bit more information than would be necessary in a scientific paper reporting on such a simple investigation. Furthermore, albeit simple, this example introduces the methodology employed by many papers~\cite{Babakol:2024:TAE,Lima:2019:HEE,oliveira2021improving,pereira2021ranking,Bonvoisin:2024:UPE}, with small variations, to compare the energy efficiency of alternative software solutions.

\subsection{Benchmark description}

The functioning of the Fannkuch Redux benchmark
\footnote{\url{https://benchmarksgame-team.pages.debian.net/benchmarksgame/description/fannkuchredux.html}} 
is easy to understand with an example. Given a number $N=4$, consider a permutation of the sequence $[1,2,3,4]$, for example, $[2, 4, 1, 3]$. The program will take the first number, 2, and reverse the order (flip) of the first 2 numbers: $[4, 2, 1, 3]$. It will repeat this procedure, taking the first element $e$ and reversing the order of the first $e$ elements until the first element becomes 1. It will do this for every permutation of $[1...N]$, where $N$ is an argument, and count the maximum number of flips among all the permutations. We use a multi-threaded Java implementation of this benchmark
\footnote{\url{https://benchmarksgame-team.pages.debian.net/benchmarksgame/program/fannkuchredux-java-1.html}}.

One useful feature of this benchmark is that its entire workload is defined by a single parameter ($N$), the size of the sequence to be processed. We execute it with its default workload consisting of one execution with $N=12$. For most benchmarks, some kind of workload will need to be defined so that it (i) is comprehensive, in terms of covering different aspects of the system under analysis, and (ii) enables comparison of different solutions or across different computing devices. The process of studying real-world scenarios to derive a workload that is repeatable, representative, and sufficiently complex is called \textit{``workload characterization''}. Jain's book~\cite{Jain:1991:ACS} includes an entire chapter on this topic. In the case of the workload we are providing to Fannkuch Redux, the idea of representativeness does not make sense and an execution of the benchmark runs for approximately 4 seconds. We briefly discuss whether this is ``sufficiently complex'' in Section~\ref{sec:collecting}.
 
\subsection{Experimental environment}

We will be running this experiment on a Macbook Air from 2020 using an Apple M1 chip (8 cores, 4 x 3.23GHz, 4 x 2.06GHz), 8GB RAM memory (LPDDR4), 8MB L3 cache, 256GB SSD, running macOS Monterey version 12.5. For a Mac, reporting this information is sufficient since there is a relatively small number of variations in terms of hardware and OS configuration. For a PC, we would provide a bit more information about the underlying hardware and the operating system, e.g., configuration for CPUFreq\footnote{\url{https://docs.kernel.org/admin-guide/pm/cpufreq.html}} on Linux machines, how many ``virtual'' processors, and, if relevant, which GPU, whether it is dedicated or not, how much memory it has, etc. During the experiments, since we are interested in measuring the energy usage when running on battery, the laptop is not connected to the power outlet and the battery is charged between 100\% and 95\%. Since mobile devices like laptops and smartphones perform aggressive dynamic voltage and frequency scaling, keeping the battery within a certain range and reporting that range is important for reproducibility. 

The benchmark is written in Java and the version installed on the machine is 21.0.2 (for both javac and the JVM). We use the default configurations for the JVM in terms of memory management, more specifically, max heap size and thread stack size are 2048MB and 1024MB, respectively\footnote{Obtained by running \texttt{java -XshowSettings:vm} and from \url{https://docs.oracle.com/en/java/javase/21/docs/specs/man/java.html}}, and the garbage colletor is G1\footnote{
    \url{https://www.oracle.com/webfolder/technetwork/tutorials/obe/java/gc01/index.html}}. 
We do not specify optimization levels for the Java compiler because it does not support them, differently from compilers for languages such as C and Rust. We run the experiments immediately after booting up the computer and have no applications running on background beyond the ones that are part of the OS.  On a Linux machine, experiments should run under similar conditions and in text mode to avoid the overhead of having a GUI. Furthermore, running in recovery mode (at least in Ubuntu-based Linux distros) is even better, since it only loads the absolutely essential for the OS to boot up\footnote{\url{https://wiki.ubuntu.com/RecoveryMode}}, which means less processes running in the background competing for scheduler time. According to the official Ubuntu wiki, \textit{``this mode just loads some basic services and drops you into command line mode.''}

\subsection{Collecting energy data}\label{sec:collecting}

\begin{figure}[tb]
    \centering
    \includegraphics[width=1\linewidth]{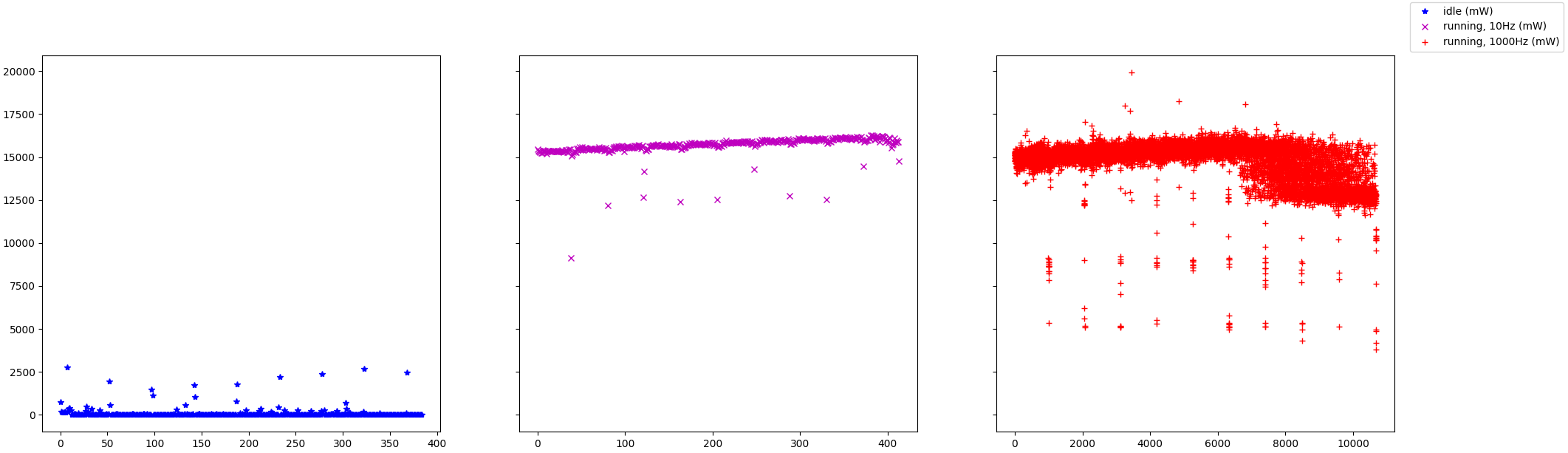}
    \caption{Power measurements (in mW) for idle (in blue), Fannkuch Redux with 10Hz sampling rate (in magenta), and Fannkuch Redux with 1kHz sampling rate (in red). }
    \label{fig:measurements}
\end{figure}

Mac systems provide a command-line utility that reports on CPU and GPU power usage called \texttt{powermetrics}. It is able to collect data at a sampling rate of up to 1kHz (one observation per milisecond). When Macs used Intel processors, \texttt{powermetrics} would leverage Intel's RAPL interface~\cite{Khan:2018:RAE} to get fairly good estimates of processor and RAM energy use. After Apple transitioned into using its own proprietary processor on Macs, the details of how \texttt{powermetrics} collects data became less clear, as Apple does not disclose them and the \texttt{man} page\footnote{\url{https://www.unix.com/man-page/osx/1/powermetrics/}} is outdated (as of \today). Notwithstanding, considering that Apple has total control over the design of its processors and operating system and many years of experience in building mobile devices, we assume that the results reported by it are at least reliable enough to be used to optimize benchmarks and applications for energy efficiency, though maybe not to compare different machines. Before we perform the experiment, we run \texttt{powermetrics} for a few seconds to gauge the idle power usage of the device. More specifically, we run the following command on the Mac terminal: 

{\scriptsize
\begin{verbatim} 
powermetrics -o output.txt -i 100 --samplers cpu_power -n 385
\end{verbatim}
}
\noindent 
This command produces output written in file \texttt{output.txt}. The sampling rate will be 10Hz (one sample every 100ms), data will pertain to the power draw of the CPU (other sources such as GPU and disks can also be specified), and overall 385 observations will be collected. The number 385 is the size of a sample taken from a population of unlimited size that provides us with a 95\% confidence level that the real value of the power is within $\pm$5\% of the observed value. After processing the data in file \texttt{output.txt}, we observe that the mean power draw of the CPU was 106.818mW, with a median of 18mW, and a standard deviation of 379.59mW. These results show that there is a lot of variation in power when the machine is idle. The left-hand plot in Figure~\ref{fig:measurements} presents the power measurements for all the collected observations. Sampling once, a few times, or even a few dozen times is not sufficient to give us confidence on the reliability of the observations, specially when gathering power, which is why we should have statistical confidence instead. In addition, some hardware components exhibit tail power states~\cite{Pathak:2011:FGP}, i.e., they stay in a high-powered state even after they stop being used, sometimes for several seconds. Examples include network adapters, sd cards, and GPS. If interacting with such components, it is essential to have buffer time between subsequent executions so that the components leave this high-powered state.

The next step is to run the benchmark whose energy usage we want to estimate. However, there is a small obstacle to that: \texttt{powermetrics} reports power at regular intervals. To use it to estimate energy, we need to \textbf{synchronize the execution} of its process with that of the benchmark we want to analyze. There are multiple ways of doing this, such as starting the \texttt{powermetrics} process directly from the benchmark's \texttt{main()} method. However, \texttt{powermetrics} requires \texttt{sudo} powers to get information about CPU, which complicates this strategy a bit. An easier approach in this scenario is to simply write a shell script that starts it in a separate process, starts the benchmark, and, when execution ends, kills powermetrics so that we know that all the data it generated is (approximately) associated with the execution of the benchmark. This script must be run with sudo powers. An example of such script is the following:

{\scriptsize
\begin{verbatim}
powermetrics -o output.txt -i 100 -s cpu_power -a  --hide-cpu-duty-cycle &
pm_id=`/bin/ps -a | grep "powermetrics" | grep -v "grep" | grep -v "su" |  awk '{print $1}'`
java  -cp . fannkuchredux 12 
sudo kill -9 $pm_id
echo "Finished execution."
\end{verbatim}
}

\noindent
This script specifies that the sampling rate is 10Hz and that \texttt{powermetrics} should not output all the information that it typically does. We modify the \texttt{main()} method of Fannkuch Redux so that it runs 10 times, instead of just one. Running \textbf{multiple times} is important because it reduces the impact of transient perturbations, e.g., due to the OS scheduler, and also helps improve cache locality and make the JVM ``hot'' by compiling parts that are executed multiple times~\cite{Pinto:2014:UEB}. In other words, it makes execution more reliable and repeatable. Ideally, we should discard the first few executions, as they may not be representative of real executions~\cite{Pinto:2014:UEB}, but in this case we will keep all of them, as it simplifies the synchronization with \texttt{powermetrics}. There is no specific number of times that is ideal to execute a benchmark. For example, when running jobs that take hours to execute, it is not beneficial to have multiple executions because transient factors become irrelevant. Additional executions are just wasted time. In general, executing a benchmark multiple times is useful to (i) guarantee that execution conditions are realistic, (ii) produce more stable estimates, and (iii) perform statistical verification. Finally, running multiple times is wasteful for work based on the premise that we should be wasting less resources, if it does not bring tangible benefits.

For convenience, we run the script preceded by the UNIX \texttt{time} command to obtain the elapsed execution time of the benchmark, as well as the processor time used to execute the process and the system overhead. Alternatively, we could also have computed execution time directly within the \texttt{main()} method, e.g., using the \texttt{System.nanoTime()} Java method invoked at the beginning and at the end of the method. This avoids the overhead of starting up the JVM, but provides less information than the \texttt{time} command. 

A relevant limitation of the employed methodology is that it only considers dynamic power. Leakage power, the power that stems from current passing through a circuit when there should be none, can also be significant~\cite{Flautner:2002:DCS}. We are not aware of any previous paper in the area of \gsd that explicitly accounts for leakage power.  

\subsection{Experiment results}\label{sec:results}

Fannkuch Redux took 42.91s to execute. The mean CPU power was 15.215W, with a median of 15.319W and a standard deviation of 0.610W. The middle plot in Figure~\ref{fig:measurements} presents the power measurements for all the collected observations, when running the benchmark.There is less variation when measuring energy footprint of a computing device under heavy load, compared to when it sits idle, if CPU frequency is kept stable. If we multiply mean power and time, we obtain the overall energy consumed by the ten executions, 652.902J. Subtracting the energy that would be consumed while the CPU is idle, we get 652.902 - (0.106W * 42.91)= 648.354. This result emphasizes the enormous difference in  energy usage of the CPU between its idle and busy states, when occupying all the CPU cores; on average, more than 140 times for this specific example. This helps us understand the idea of racing to idle~\cite{Albers:2012:RIN,Fu:2018:RIN} to save energy. Dividing the obtained result by 10 (the number of executions) tells us how much the benchmark consumes in one execution: \textbf{64.835J}, answering our research question.

Running tools such as \texttt{powermetrics} or NVIDIA's \texttt{smi}\footnote{https://developer.nvidia.com/system-management-interface} creates a conundrum. These tools report power and thus with a higher sampling rate, we can potentially obtain more accurate data, as discussed in Section~\ref{sec:measure}. However, they run in their own processes and thus consume resources themselves. Furthermore, since they must somehow report the results, I/O operations are required, e.g., writing to a file, which creates delays for the sampling process. Therefore, if the sampling rate is high, this may have an impact on the results. For example, running the same script providing \texttt{-i 1} to \texttt{powermetrics}, i.e., 1kHz or one thousand observations per second, execution takes 46.79s and the mean power is 14.740W, with a median of 14.838W and a standard deviation of 0.899W. The right-hand plot in Figure~\ref{fig:measurements} presents the power measurements running Fannkuch Redux while gathering data with this higher sampling rate. The plot and the numbers show that the higher sampling rate reduces the mean power draw of the CPU. This is intuitive: the scheduler must switch more often between the benchmark and \texttt{powermetrics} and these switches produce more frequent periods of lower power. Discounting the energy consumed while idle as well, the CPU consumed 684.754J, 5.6\% more than for a sampling rate of 10Hz. The higher sampling rate imposed a non-negligible overhead. Comparing the samples of power readings for the the two cases, 10Hz and 1,000Hz, using Mann-Whitney's U test yields $p < 10^{-10}$, with a large effect size (Cliff's Delta) of 0.894. The number of collected observations illustrates this difference. On the one hand, a sampling rate of 10Hz produced 406 data points, less than the expected 429 (one every 100ms, during 42.91s) but not by much. On the other hand, a sampling rate of 1kHz produced 10019 observations, i.e., less than $\frac{1}{4}$ of the 46,000 we should be getting. 

This example does not aim to advocate the use of any specific sampling rate. Instead, it is necessary to study the overhead of the energy measurement tools so as to avoid overhead that may skew the results. Sometimes this involves using different tools. For example, in our experience, NVIDIA smi imposes non-negligible overhead even for relatively low sampling rates, e.g., 5Hz. We have observed that switching to the NVIDIA Management Library\footnote{\url{https://developer.nvidia.com/management-library-nvml}} (NVML), or some wrapper\footnote{\url{https://pypi.org/project/pynvml/}} around it, yields significantly better results. 

In summary, the challenge is to strike a balance between a high enough sampling rate and obtaining accurate and representative data, with a low overhead. 

\section{Approaches to estimate energy usage}\label{sec:approaches}

This section provides an overview to some approaches to estimate software energy consumption. Section~\ref{sec:counters} introduces the combination of  hardware performance counters and software. This is arguably the easiest solution to estimate (measure) energy usage in contemporary hardware, in particular for laptops and servers, with sufficient accuracy~\cite{Khan:2018:RAE} to compare alternative software solutions. Section~\ref{sec:external} then dips a foot into external measurement equipment, which has the advantage of providing potentially precise, non-intrusive measurements, at the cost of additional work for setting up and synching up measurement and execution. Finally, Section~\ref{sec:models} glosses over modeling to estimate energy consumption, focusing specifically on one strategy adopted by multiple papers. 

\subsection{Hardware performance counters and software}\label{sec:counters}

This approach, as the name of this section suggests, involves a combination of hardware performance counters to keep track of data about energy footprint and software able to collect that data. The performance counters are registers that are regularly updated with information such as the number of joules consumed since the processor was started, the current power level of the processor(s), the energy consumption of the RAM, and current temperature. The software part can be a proprietary tool such as NVIDIA smi, system tools such as \texttt{perf}, or a custom built application that directly accesses the performance counters. For this to be possible, hardware manufacturers make available interfaces through which tools can obtain the data. 

The main representative of this approach is Intel's Running Average Power Limit (RAPL), available for its line of processors for more than ten years (as of June 2024), since the Sandy Bridge architecture. RAPL can obtain data about energy footprint and also limit (cap) the average power draw of components of the processor, although we do not explore this feature here. Data is made available at a high sampling rate of 1kHz and imposes a low overhead when we only collect data about energy usage~\cite{Huang:2015:MCH}. Previous work has shown that the data that RAPL produces is very similar to what can be obtained with external, high frequency measurement equipment~\cite{Khan:2018:RAE}. This combination of ease of use, accuracy, and the ability to collect data specifically about the energy footprint of the processor and neighboring components makes RAPL popular with researchers and developers. To use RAPL, \texttt{sudo} powers are required, just like with \texttt{powermetrics}. 

RAPL reports energy usage of different \textit{power domains}, physical hardware elements for which it is possible to manage power. Examples of such hardware elements include the processors (PP0), RAM (DRAM), uncore components (PP1 -- L3 cache, memory controllers), and package (PKG -- the cores plus uncore and on-chip interfaces for memory and IO). Not every power domain is supported by every Intel architecture and not even every device with a given architecture, e.g., on-socket GPUs are only supported for desktop models~\cite{Khan:2018:RAE}. Data about the energy consumption of different power domains is kept at model-specific registers (MSRs). As mentioned before, they can be accessed directly or by means of tools such as \texttt{perf} and the Intel Performance Counter Monitor\footnote{\url{https://github.com/intel/pcm}}. If performing direct accesses, two important factors must be accounted for. First, different units are reported, based on the architecture and even the power domain. According to Khan et al.~\cite{Khan:2018:RAE}, 

\begin{quote}
    \textit{``[...] energy is counted in multiples of model-specific energy units. Sandy Bridge uses energy units of 15.3 microjoules, whereas Haswell and Skylake uses units of 61 microjoules. In some CPU architectures such as Haswell-EP, DRAM units differ from CPU energy units. The units can be read from specific MSRs before doing energy calculations. ''} 
\end{quote}

\noindent
Second, MSRs only have 32 bits. Since they report at a granularity of tens of microjoules  (1J = 1,000mJ = 1,000,000$\mu$J), they can wrap around fast. A processor with a power draw of 80W\footnote
{\url{https://www.intel.com/content/dam/doc/white-paper/resources-xeon-measuring-processor-power-paper.pdf}} 
at full power would cause the corresponding MSR to wrap around in approximately 13 minutes on Sandy Bridge. Existing tools and libraries that leverage RAPL, such as Scaphandre\footnote{\url{https://github.com/hubblo-org/scaphandre}} and PowerJoular~\cite{Noureddine:2022:PJM}, account for these issues and are able to consistently report energy usage estimates in joules (and sometimes do even more!).

One last thing to observe about RAPL is that Intel introduced in late 2020 a feature in its line of processors called \textit{``energy filtering''}, associated with its Software Guard
Extensions\footnote{\url{https://www.intel.com/content/www/us/en/developer/tools/software-guard-extensions/overview.html}} 
(SGX), for security reasons~\cite{Intel:2020:RAP}. This feature, when enabled, causes distortion in the results reported by RAPL. This phenomenon has been reported by Intel~\cite{Intel:2020:RAP} and observed experimentally by independent researchers~\cite{Tarara:2022:RSE,Schone:2024:EEF}. Both Tarara and Sch\"one et al. report that distortions happen specially in idle or low-power states. On top of this, Sch\"one and colleagues show that the sampling interval is about 8ms for the PP0 domain (processor only) when filtering is on, although it is still 1ms for the PKG domain. Therefore, this is another aspect that, if possible, should be reported in experiments in Intel machines that leverage RAPL.

\begin{figure}[tb]
    \centering
    \includegraphics[width=0.8\linewidth]{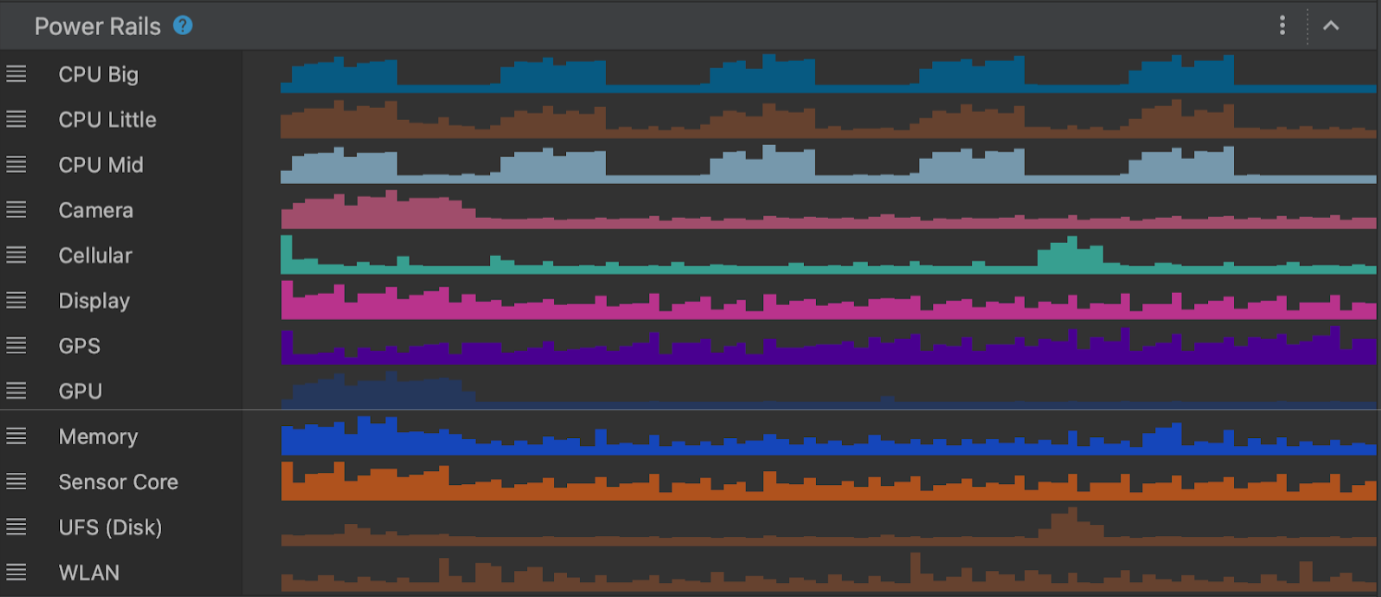}
    \caption{Android Power Monitor presenting the power draw of different power rails of a Pixel 6 smartphone. Source: \url{https://developer.android.com/studio/profile/power-profiler}}
    \label{fig:powerrails}
\end{figure}

This approach of making information about energy usage available by means of software tools that read hardware counters is also available in other platforms. For example, AMD offers an interface\footnote{\url{https://github.com/amd/amd_energy}} similar to RAPL, including the use of power domains, MSRs, and accessible through tools such as \texttt{perf}. Furthermore, the power draw of NVIDIA GPUs can be inspected by using the NVIDIA smi tool, which reports with a sampling rate of up to 1kHz and works very similarly to Apple's \texttt{powermetrics}. The latter is another example of the ``software'' side of this approach. In this case, as with NVIDIA smi, we cannot directly access the performance counters. One thing to bear in mind is that we have observed, at least anecdotally, that smi imposes a significant overhead even at relatively low sampling  rates, e.g., 10Hz. As mentioned in Section~\ref{sec:results}, to reduce that overhead, one can instead use the NVIDIA Management Library (NVML), at the cost of having a bit more implementation work. More recently, Pixel 6 devices running at least Android 10 offer the On-Device Power Rails Monitor (ODPM). It organizes the data about the power draw of different components in a smartphone in terms of \textit{``power rails''}. The data is accessible through Android's Power Profiler\footnote{\url{https://developer.android.com/studio/profile/power-profiler}}. Figure~\ref{fig:powerrails} presents a screenshot of the Power Profiler, with power usage data for twelve rails, including different processors (``Big'', ``Little'', ``Mid''), cellular network, and display.2

\subsection{Specialized measurement hardware}\label{sec:external}

Using specialized measurement hardware presents some advantages compared to leveraging performance counters and software to access them. As discussed before, measurement hardware is non-intrusive in the sense that the measurement process does not require changes to the program or to its execution environment, nor is there any runtime overhead stemming from the measurement process. Furthermore, it can be more precise than alternatives such as RAPL, with the possibility of reaching sampling rates of 5kHz or more, if one is willing to spend the money to acquire high-end equipment. 

Measurement hardware also has disadvantages. Synchronizing benchmark execution with the energy measurement process is more difficult, often requiring either an additional device to start up both the benchmark and the measurement procedure in tandem or some a posteriori alignment based on timestamps and linear regression. Furthermore, there are two situations where the use of measurement equipment may require some additional setup: when measuring the energy footprint of software running on battery-based devices and measuring the energy usage of specific parts of a computing device. In ths first case, as illustrated in Figure~\ref{fig:power}, the battery must be removed to avoid misrepresentation of the actual power draw, due to the battery charging process, and the dynamic frequency and voltage scaling that happens while the device is running on battery power. If the battery is removed, some kind of power source must be connected to the device. In addition, contemporary phones require the battery to be in place to work, i.e., the phone performs some validation with the battery and does not function without it. In this case, it is necessary to tinker with the part of the battery responsible for the interface between the battery cells and the device itself~\cite{cruz2021tools}. This was required for the setup in Figure~\ref{fig:power}. 


\begin{figure}
    \centering
    \begin{minipage}[t]{.5\textwidth}
      \centering
      \includegraphics[width=.7\linewidth]{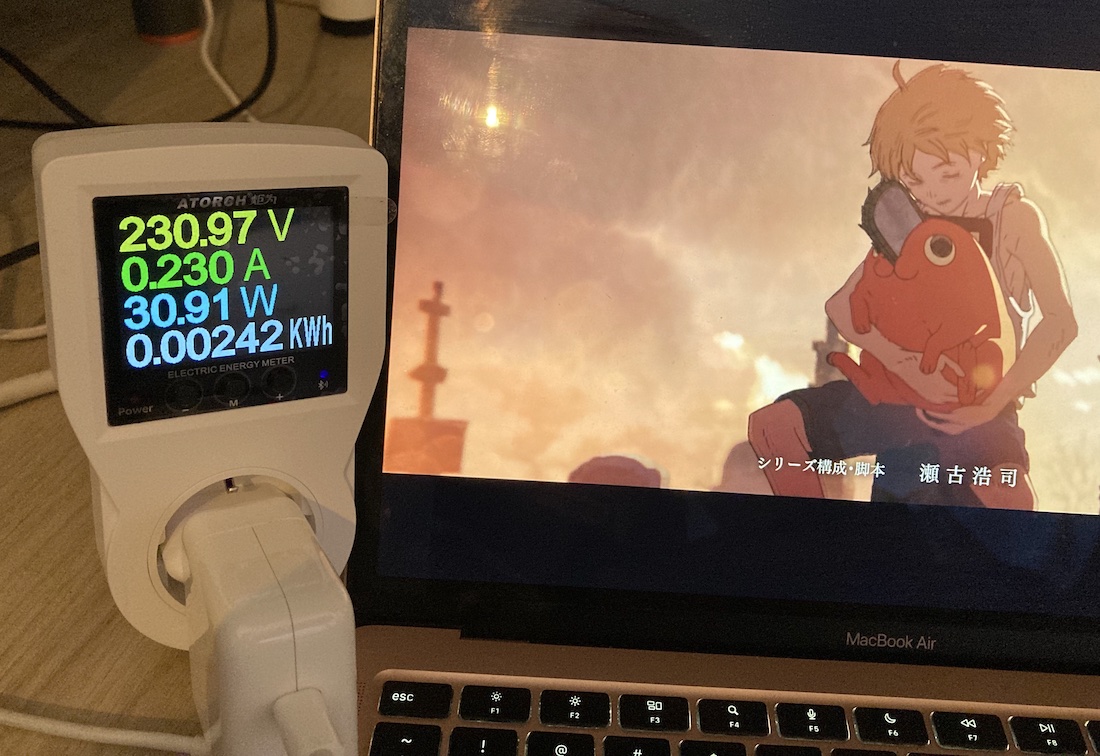}\\(a)
    \end{minipage}%
    \hspace*{0.3cm}
    \begin{minipage}[t]{.5\textwidth}
      \centering
      \includegraphics[width=.6\linewidth]{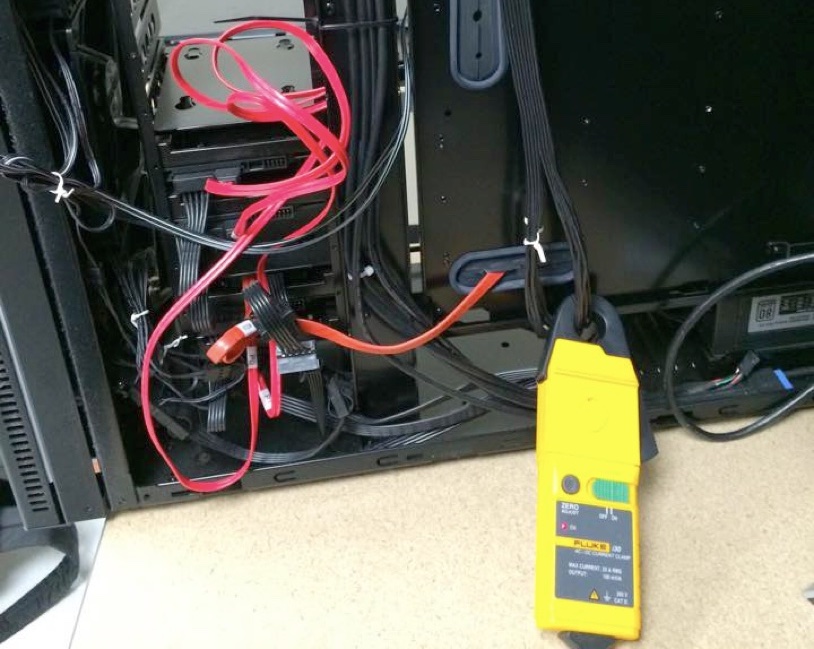}\\(b)
    \end{minipage}
    \caption{Two measurement setups with external measurement equipment: (a) a power meter placed between the power outlet and the power plug capturing the power draw of a notebook, and (b) a current clamp capturing the current flowing to the processor of a server machine.}
    \label{fig:external}
  \end{figure}

Some power meters can be placed between the power outlet and the power plug. If the device under measurement works even with the battery removed, is not battery powered, or can have the charging turned off for measurement purposes, this solution can save significant work. Figure~\ref{fig:external}(a) presents an example of this approach. In the figure, the laptop's power adapter is connected to the meter which is connected to a power outlet. The display shows voltage, current, power, and energy. Although the laptop is battery-powered, this one can stop charging the battery when its charge level is high enough and it is connected to a power source. 

Yet another solution is to use equipment to capture the current flowing through wires. If the voltage is also known or can be determined (e.g., by using an oscilloscope), we can calculate the power draw as the product of current and voltage. This approach can be used to measure both whole-system energy usage and the energy usage of specific parts, if they are connected to a power source by a wire or cable. Figure~\ref{fig:external}(b) shows an example of current clamp (yellow, in the right-hand side of the figure) connected to the cable providing power to the processor of a server machine. 

To be used in the kind of experiment presented in Section~\ref{sec:ex}, measurement equipment must support telemetry. This enables us to gather the collected data in a format that can be easily analyzed by a computer. Most high end solutions provide those capabilities. They are not universal and in some cases additional equipment is required. On the one hand, the meter in Figure~\ref{fig:external}(a), for example, does not provide the ability to easily export the readings. On the other hand, a current clamp such as the one in Figure~\ref{fig:external}(b) requires additional equipment such as an oscilloscope or an analog to digital converter, if we want to be able to process the results using a computer.

\subsection{Analytical models}\label{sec:models}

Modeling is a complex topic and we have no hope of even scratching the surface in this document. Models can vary widely in terms of complexity and  granularity. For example, a state-of-the-art analytical model capable of estimating GPU power, such as AccelWattch~\cite{Kandiah:2021:APM}, can consider a wide range of factors, including \textit{``constant, static, and dynamic power consumption in the presence of DVFS, thread divergence, intra-warp functional unit overlap, variable SM [Streaming Multiprocessor] occupancy, and power gating.''}. Conversely, to model the energy footprint of the execution of a GPU heavy application, we can simply multiply the rate of floating operations per second (FLOPS) of the underlying GPU by its execution time (more on this later). Other factors to consider when building a model for energy include  the employed mathematical framework, required data, moment (predictive, on-line, or a posterior estimation), and produced output.

In this section we focus on a specific type of modeling solution. The idea is to enable developers to compare alternative software solutions without directly considering the specific number of joules that each one requires under a certain workload. This type of solution has appeared multiple times in the literature~\cite{Hao:2013:EMA,oliveira2021improving,Pathak:2011:FGP} and is easy to use or even build from scratch. In addition, software developers are not used to thinking in terms of joules, with lack of appropriate KPIs being cited as a major obstacle to taking energy efficiency into account in software development~\cite{Ournani:2020:REC}. On top of this, hardware platforms can impact energy consumption differently~\cite{Jacques2024battery,oliveira2021improving}. Thus, placing the emphasis on the ability to compare results for software is a practical approach that circumvents these issues.
 
To the best of our knowledge, the template for models of how programs consume energy, from a software developer's perspective, was devised by Pathak et al.~\cite{Pathak:2011:FGP} and refined by Hao et al.~\cite{Hao:2013:EMA}. The basic idea is to have two models: (i) a platform-independent, application-dependent model that estimates how intensively the system under analysis performs operations of interest, and a (ii) platform-dependent, application-independent model that estimates the energy footprint of individual operations. The term \textit{``operation''} is used in a loose sense: it can refer to both machine-level instructions and higher level concepts such as data structure methods and system calls. The combination of these two models produces an estimate of the energy footprint of the software system (or some proxy metric) that can be used to compare different ways in which it can be built or executed. 

More concretely, the first model is typically built by leveraging static~\cite{Hao:2013:EMA,oliveira2021improving} or dynamic~\cite{Chouwdury:2015:SCB,Hasan:2016:EPJ,Pathak:2011:FGP} analysis. The second model can be obtained by experimental measurement of the cost of individual operations~\cite{Hao:2013:EMA,oliveira2021improving}, estimation based on power states~\cite{Pathak:2011:FGP}, or type of instruction.~\cite{Chouwdury:2015:SCB}. On top of these concerns, specific models also account for issues such as path-dependence~\cite{Hao:2013:EMA} and whether they are thread-safe or not~\cite{oliveira2021improving}. 

This modeling approach can be applied in a non-obvious but simple manner to systems that employ deep learning and large language models. In this scenario, the ``operation'' of interest can be all floating point operations, which are usually executed by GPUs. We can estimate the number of floating point operations required to perform a certain task. The rate of FLOPS that contemporary GPUs can perform is publicly disclosed by their 
manufacturers\footnote{\url{https://www.nvidia.com/en-us/geforce/graphics-cards/40-series/rtx-4080-family/}}. 
GPUs are responsible for the bulk of the work to train and use deep learning systems. Since that work consists of performing a large number of floating point operations, measuring the time spent in these tasks and multiplying it by the number of FLOPS of the underlying GPU gives us an approximate measure that we can use to compare machine learning solutions, in terms of their energy footprint. If the execution environment is kept unchanged, this is a reasonable metric~\cite{Desislavov:2023:TAI}. If factors such as the hardware configuration, deep learning framework or library~\cite{georgiou2022green,Stojkovic:2024:TGL}, runtime system~\cite{Alizadeh:2024:PES}, and machine learning task~\cite{Alizadeh:2024:PES,Jacques2024battery,georgiou2022green} can vary across executions, results may differ significantly. 







\section{Some observations about tools}\label{sec:tools}

The goal of this document is not to talk about tools to gather data about energy efficiency (``measurement tools'', for simplicity). Cruz and his collaborators~\cite{cruz2021tools} wrote a nice post about this topic. It is a bit outdated, e.g., it does not talk about  (as of \today) solutions such as the Snapdragon Profiler, which replaced the now-defunct Trepn)\footnote{https://www.qualcomm.com/developer/software/snapdragon-profiler}, nor the On-Device Power Rails Monitor available in recent Android smartphones, and talks about discontinued solutions, such as Intel Power Gadget. Nevertheless, it is still a useful resource if the reader is interested in a hands-on perspective, written by folks who have experience in using these tools. 

Even though discussing tools is outside the scope of the document, it is useful to make some considerations with the aim of promoting more rigorous data collection. The employed approach to collect energy usage data should be (i) transparent, (ii) automatic, and (iii) precise. \textbf{Transparency} means that it is always clear what is being reported by a measurement tool, i.e., the tool does not try to hide the hurdles it encounters in ways that may compromise the reliability of the reported results. A good example of a tool that I believe could be more transparent is CodeCarbon\footnote{\url{https://codecarbon.io/}}. It's a well-known, mature tool that made a great job popularizing energy measurement and making it accessible to many energy-conscious software developers. It includes documentation\footnote{https://docs.codecarbon.io/latest/explanation/methodology/?h=constant} about how it collects the data that it reports. This documentation is not always clear, though. For example, it states that \textit{``If you do not want to give sudo rights to your user, then CodeCarbon will fall back to constant mode to measure CPU energy consumption''}. The problem is that I could not find\footnote{Emphasis on ``I''. Maybe the info is there. I can say that I spent some time searching. Also, the information used to be easy to find before the CodeCarbon page changed.} what ``constant mode'' means. The attempt to hide the potential complications of collecting energy data may be misleading. I've met fellow developers who used CodeCarbon and weren't even aware that the energy data they were getting was not accessing the machine's underlying MSRs due to lack of \texttt{sudo} powers. 

Measurement tools also need to be automatic: it should be possible to write scripts to automatically use and manage them. Software-based tools usually meet this requirement, either directly, i.e., they are usable via language-specific libraries, e.g., pyNVML, which provides programmatic access to power drawn by NVIDIA GPUs, or indirectly, i.e., they are command-line tools whose outputs can be processed by shell-based scripts, e.g., \texttt{perf} and \texttt{nvidia-smi}. For hardware-based meters, this implies that telemetry support is essential. I have a cheap meter (Figure~\ref{fig:external}(a)) that can be connected directly to the power outlet to measure system level power draw. It is interesting to observe the results and easy to use, but it's not useful for the kind of study we do. First, because it does not support exporting the collected data. Second, because it has a very low sampling rate of 1Hz, which takes me to the last aspect, \textbf{precision}. 

Precision in this context is about two aspects: (i) reporting resource usage that matches what is actually being used and (ii) doing so at granularity that enables us to make conclusions about the impact of interventions. Hardware monitors that do whole-system measurement are particularly good at this. These instruments calculate power as the product of current and voltage and therefore there is little margin for error. We can also leverage software-based approaches that collect data from hardware-based performance counters. This is what RAPL and derived tools like \texttt{perf}, pyRAPL, PAPI, Scaphandre\footnote{https://github.com/hubblo-org/scaphandre}, PowerJoular\footnote{https://github.com/joular/powerjoular} and others do. These tools have a reasonable sampling rate close to 1KHz and there is reliable evidence that what they report is very close to what one would get by using a hardware monitor~\cite{Khan:2018:RAE,Tarara:2022:RSE}. Some tools do not clarify how they make their estimates of power usage, e.g., Apple's \texttt{powermetrics} and NVIDIA's \texttt{nvidia-smi}. We trust these tools implicitly because they are developed by the companies that design the hardware they aim to monitor.  Other tools clearly document the compromises they make to estimate when there is no information available. For example, CodeCarbon estimates RAM power use as a flat wattage per memory module (DIMM), which changes per platform. This is well-documented, but it is likely not precise enough to be useful.

\section{Further study}\label{sec:further}

Besides the many documents we refer to throughout the paper, in this section we provide a few additional pointers. John~\cite{John:2000:PET} has written a succinct text on performance evaluation of computing systems with a specific focus on the distinction between measurement and modeling, as well as the various quality attributes that can be considered for a performance evaluation approach. For an in-depth treatment of the topic, Raj Jain's book~\cite{Jain:1991:ACS} is still useful, in spite of the age. 

Vincent Weaver used to maintain a homepage titled ``Reading RAPL energy measurements from Linux''~\cite{weaver:RAPL}. It is an invaluable resource if one wants to access RAPL data in a more direct manner without going through the manuals of the processors. In particular, it presents two tables indicating what information is available for each family of processors, considering both Intel and AMD, and whether they are supported by Linux tools such as \texttt{perf} and PAPI. The page seems to not be under active maintenance anymore, unfortunately.

There are factors that may impact the reliability of experiments aiming to estimate energy consumption beyond the ones discussed in Section~\ref{sec:ex}, such as the use of Turbo Boost. Ournani and colleagues~\cite{Ournani:2020:TEC} perform an empirical evaluation aiming to identify and quantify some of them. 

In a similar vein, the Green Coding website\footnote{https://www.green-coding.io/} includes interesting, hands-on articles related to the energy efficiency of software. They cover diverse topics such as the impact of SGX and energy filtering, the usefulness of CPU utilization as a metric for resource usage, and the carbon cost of downloading Zoom. Really good stuff. 

\section{Acknowledgments}

We would like to thank Ivano Malavolta and Negar Alizadeh for kindly reading and providing feedback on a preliminary version of this document. We would also like to thank Arne Tarara for writing very interesting articles on low level aspects of \gsd~ and, more specifically, for an enlightening discussion about SGX and energy filtering in Intel machines. 

\begin{small}
\bibliographystyle{plain}
\bibliography{references}
\end{small}

\end{document}